\documentclass[a4paper, fleqn]{cas-sc}

\usepackage[authoryear]{natbib}
\usepackage{lipsum}
\usepackage{mathtools, amssymb, amsmath,amsthm}
\usepackage{float} 
\usepackage{bookmark}
\usepackage{xcolor}
\allowdisplaybreaks
 
\def\bs{\boldsymbol}
\def\bf{\mathbf}
\def\R{\mathbb{R}}
\newcommand{\indep}{\perp \!\!\! \perp}
\newtheorem{thm}{Theorem}
\newtheorem*{thm*}{Theorem}

\newtheorem{lemma}[thm]{Lemma}
\let\printorcid\relax 

\begin{document}

\let\WriteBookmarks\relax
\def\floatpagepagefraction{1}
\def\textpagefraction{.001}
\shorttitle{Stabilized Sufficient Dimension Reduction}
\shortauthors{Boonstra et~al.}

\title[mode = title]{Precision Matrix Regularization in Sufficient Dimension Reduction for Improved Quadratic Discriminant Classification}   

\author[label1]{Derik T. Boonstra}
\author[label1]{Rakheon Kim}
\author[label1]{Dean M. Young}
 
\address[label1]{Department of Statistical Science, Baylor University, Waco, TX 76798-7140}

\begin{abstract}
Sufficient dimension reduction (SDR) methods, which often rely on class precision matrices, are widely used in supervised statistical classification problems. However, when class-specific sample sizes are small relative to the original feature-space dimension, precision matrix estimation becomes unstable and, as a result, increases the variability of the linear dimension reduction (LDR) matrix. Ultimately, this fact causes suboptimal supervised classification. To address this problem, we develop a multiclass and distribution-free SDR method, stabilized SDR (SSDR), that employs user-specified precision matrix shrinkage estimators to stabilize the LDR projection matrix and supervised classifier. We establish this technique with the theoretical guarantee of preserving all classification information under the quadratic discriminant analysis (QDA) decision rule. We evaluate multiple precision matrix shrinkage estimators within our proposed SSDR framework through Monte Carlo simulations and applications to real datasets. Our empirical results demonstrate the efficacy of the SSDR method, which generally improves classification accuracy and frequently outperforms several well-established competing SDR methods.
\end{abstract}

\begin{keywords}
Central subspace \sep Covariance sparsity \sep Heteroscedasticity \sep Shrinkage estimator \sep Singular value decomposition 
\end{keywords}

\maketitle

\setlength{\belowdisplayskip}{4pt} \setlength{\belowdisplayshortskip}{4pt}
\setlength{\abovedisplayskip}{4pt} \setlength{\abovedisplayshortskip}{4pt}

\section{Introduction}\label{sec:intro}

For statistical discriminant analysis and other parametric multivariate statistical methods, one generally must estimate the precision matrix. However, this task becomes challenging when the feature-space dimension, $p$, is large relative to the class-specific sample size, $n_{i}$, $i = 1, \ldots, k$, where $k$ is the \textit{a priori} number of classes. With advancements in computing, recent literature has increasingly focused on addressing dimensionality issues in precision matrices and classification in high-dimensional scenarios, $n_{i} < p$. However, issues may still arise in discriminant analysis for heteroscedastic populations due to precision estimation when $n_{i} > p$, specifically, when $p < n_{i} < p^{2}/2$, which \cite{bellman1961} first referred to as \textit{the curse of dimensionality}. In general statistical methods and discriminant analysis, we can avoid the \textit{curse of dimensionality} by using 
\textit{SDR} to reduce the feature-space dimension to a lower dimension, $q < p$.

\cite{Anderson1951} developed \textit{QDA} as a generalization of \cite{fisher1936}'s \textit{linear discriminant analysis} (\textit{LDA}) to incorporate differences in covariance matrices. However, \cite{Gaynanova2016} found that \textit{QDA} may suffer from the \textit{curse of dimensionality} and perform poorly in terms of the \textit{conditional error rate} (\textit{CER}) if an insufficient sample size is used to estimate the $(k-1) + kp + kp(p+1)/2$ parameters required under heteroscedasticity. As a result, they concluded that even when covariance matrices differ, \textit{QDA} can perform worse than \textit{LDA} in terms of the \textit{CER} when training-sample sizes are small. This result occurs because pooling covariance matrices in \textit{LDA} substantially reduces the total number of parameters to estimate. Alternatively, by projecting the data onto a lower dimensional  and information preserving subspace, \textit{SDR} techniques reduce the number of free parameters and allow for the additional benefit of preserving heteroscedastic classification information under the \textit{QDA} decision rule. More specifically, let $Y \in \R$ be the univariate response of $\bf{X} \in \R^{p}$. Then, \textit{SDR} techniques aim to identify the smallest possible reduced data dimension $q < p$ such that $\bf{R}(\bf{X}) \in \R^{q}$  where no loss of information occurs in $Y|\bf{R}(\bf{X})$ with respect to $Y|\bf{X}$.  

In \textit{SDR}, the dimension reduction is usually constrained to be a linear transformation such that $\bf{R}(\bf{X}) = \bf{B}^{T}\bf{X}$, $\bf{B} \in \R^{p \times q}$. Thus, we say the dimension reduction is a \textit{sufficient linear reduction} if it satisfies at least one of the following: (i) $Y \indep \bf{X}|\bf{B}^{T} \bf{X}$, (ii) $\bf{X}|(Y,\bf{B}^{T}\bf{X}) \sim \bf{X}|\bf{B}^{T}\bf{X}$, or (iii) $Y|\bf{X} \sim Y|\bf{B}^{T}\bf{X}$. The subspace $\mathcal{S}$, defined by $\text{span}(\bf{B}) \subseteq \R^{p}$, is then called a \textit{dimension reduction subspace} (\textit{DRS}). Moreover, $\mathcal{S}$ is said to be a \textit{minimum DRS} for $Y|\bf{X}$ if $\dim(\mathcal{S}) \leq \dim(\mathcal{S}_{DRS})$ for every \textit{DRS}, $\mathcal{S}_{DRS}$. However, a \textit{minimum DRS} may not be unique. To address this issue, \cite{cook1998} first introduced the \textit{central subspace} (\textit{CS}), defined as the intersection of all dimension reduction subspaces, denoted as $\mathcal{S}_{Y|\bf{X}} = \cap \mathcal{S}_{DRS}$. Given that $\cap \mathcal{S}_{DRS}$ itself is a \textit{DRS}, then under mild conditions given by \cite{cook1998}, $\mathcal{S}_{Y|\bf{X}}$ exists and is the \textit{unique minimum DRS}, serving as the target subspace for most \textit{SDR} methods. Throughout the paper, we assume the existence of the \textit{CS}. 

For supervised classification, most \textit{SDR} methods are used in a two-stage reduce-and-classify approach. That is, we first derive an \textit{LDR} projection matrix to reduce the feature-space dimension and then apply an appropriate supervised classification rule. These \textit{SDR} methods often construct their \textit{LDR} projection matrix with some function of the first two conditional moments for each labeled population. This \textit{SDR} approach includes the popular \textit{Sliced Inverse Regression (SIR)}, proposed by \cite{li1991}, which reduces the dimensionality by estimating 
subspaces that capture the response variability by slicing the data along different directions. To enhance robustness, \cite{cook1991} extended \textit{SIR} by incorporating the variability within each slice through the $k$ conditional covariance matrices. Other \textit{SDR} techniques construct a targeted \textit{DRS} by optimizing  differentiable multivariate functions. For example, \cite{zhang2019} proposed a envelope discrimination method, which estimates reduced subspaces and uses envelope methods to develop a classification rule within the same model.  Alternative  \textit{SDR} methods have also been proposed, such as those by \cite{cook_and_zhang2014}, \cite{lin2019}, \cite{qian2019}, \cite{wu2022}, \cite{zeng2024}, and \cite{mai2025}. For a comprehensive review of \textit{SDR}, see \cite{cook2018} and \cite{li2018}. 

Here, we propose and evaluate a new \textit{SDR} technique, \textit{Stabilized SDR} (\textit{SSDR}), which utilizes a user-specified precision matrix shrinkage estimator to stabilize the \textit{LDR}  matrix when class parameters must be estimated. By preserving the differences among covariance matrices and reducing variability by regularizing the \textit{k} precision matrices, our proposed \textit{SSDR} method maintains or improves upon the full-feature estimated \textit{CER} in smaller data dimensions under the \textit{QDA} decision rule. Moreover, we have found strong evidence that the estimated \textit{CER} is often significantly reduced in cases when the class-specific training-sample sizes are small relative to the original feature-space dimensionality. Using \textit{Monte Carlo} (\textit{MC}) simulations and real datasets that vary in training-sample sizes, feature-space dimensions, and number of classes, we contrast the efficacy of four precision matrix shrinkage estimators that can be utilized with this new \textit{SDR} method.  For a comprehensive review of other precision matrix estimators, see \cite{fan2016}. Additionally, we demonstrate that our proposed \textit{SSDR} method provides superior classification performance in terms of smaller estimated \textit{CER}s by frequently outperforming competitive \textit{SDR} and supervised classification techniques.
 
The remainder of the paper proceeds as follows. In Section \ref{sec:main}, we establish the notation used throughout the paper and formally define the \textit{QDA} classifier. Next, we derive a theoretical \textit{SDR} method that preserves the optimal error rate under the \textit{QDA} decision rule when all group parameters are known. For the usual case when class parameters must be estimated, we provide a modification that yields improved \textit{LDR} projection matrix stabilization by employing a shrinkage estimator for each class-specific precision matrix. In Section \ref{sec:estimators}, we introduce four precision matrix estimators for possible use in our proposed \textit{SSDR} technique. We next describe our \textit{MC} simulation design for multiple population configurations and present simulation results in Section \ref{sec:mc_sim}. In Section \ref{sec:real_data}, we describe our application to several real datasets to further assess the utility of our proposed \textit{SSDR} method and then discuss the results. Moreover, we demonstrate that, in most of the real data applications considered, our proposed \textit{SSDR} method outperforms several current competing \textit{SDR} and classification techniques in terms of the estimated \textit{CER}. Lastly, in Section \ref{sec: discussion}, we discuss our new SDR methodology and results.

\section{Notation, Preliminaries, and an SDR Method}\label{sec:main}

\noindent The following notation will be used throughout the remainder of the paper. Let $\R^{m \times n}$ represent the set of all $m \times n$ real matrices. For $i = 1, \ldots, k$, let $\Pi_i$ represent the $i^{th}$  distinct population or class, let $\pi_i$ be the \textit{a priori} class membership,  let $\bs\Sigma_{i} \in \R^{p\times p}$  denote the $i^{th}$ population covariance matrix, and let $\bs\mu_{i} \in \R^{p}$ be the population mean vector for the $i^{th}$ population. Also, let $\mathbb{S}^{p} \subset \R^{p \times p}$ represent  the set of $p \times p$ real symmetric matrices, and let $\mathbb{S}^{p}_{+} \subset \mathbb{S}^p$ denote the interior of the cone of $p \times p$ real symmetric positive-definite matrices. For  $\bf{A}\in\R^{m\times n}$, we let $\bf{A}^+ \in \R^{n \times m}$ denote the Moore-Penrose pseudo-inverse, $\mathcal{C}(\bf{A})$ represent the column space, and $\mathcal{N}(\bf{A})$ denote the null space of $\bf{A}$. Lastly, let 
$\bf{A} = \bf{UDV}^T$ be the  \textit{singular value decomposition} (\textit{SVD}) of \cite{eckart1936}, where $\bf{U} \in \R^{m \times m}$  and $\bf{V} \in \R^{n \times n}$ are orthonormal matrices, and $\bf{D} \in \R^{m \times n}$ is a diagonal matrix of the singular values of $\bf{A}$. 

\subsection{Quadratic Discriminant Analysis}
\noindent Suppose one desires  to obtain a supervised classification decision rule that partitions the feature-space into $k$ disjoint regions, $\R_{\Pi_{i}}$, $i = 1, \ldots, k$, corresponding to each respective population. Then, we assign an unlabeled observation, $\bf{x}\in \R^{p}$, to $\Pi_{i}$ if $\bf{x} \in \R_{\Pi_{i}}$.  We consider the criterion function for the $i^{th}$ class to be
\begin{equation}\label{eqn:d_i}
d_i(\bf{x}) \coloneqq \log\left| \bs\Sigma_i \right| - 2 \log\left(\pi_i\right) + \left(\bf{x} - \bs\mu_i\right)^T\bs\Sigma^{-1}_i\left(\bf{x} - \bs\mu_i\right), \quad i = 1, \ldots, k.
\end{equation}
Thus, we obtain the well-established \textit{QDA} decision criterion. That is, \textit{QDA} classifies an unlabeled observation $\bf{x}$ into the class $\Pi_i$ if $d_i(\bf{x}) = D(\bf{x})$, where
\begin{equation}  \label{eqn:QDA}
D(\bf{x}) \coloneqq \min\left\{d_i(\bf{x}): i = 1, \ldots, k \right\}. 
\end{equation}
Generally, we must estimate the population parameters $\bs\mu_i$ and $\bs\Sigma_i$ 
from the training data such that $n_i > p$. Thus,  \eqref{eqn:d_i} becomes 
\begin{equation} \label{eq:d_hat}
\widehat{d}_i(\bf{x}) \coloneqq \log\left| \bf{S}_i \right| - 2 \log\left(\pi_i\right) + \left(\bf{x}-\bar{\bf{x}}_i\right)^T\bf{S}^{-1}_i\left(\bf{x} - \bar{\bf{x}}_i\right), \quad i = 1, \ldots, k,
\end{equation}
where $\bar{\bf{x}}_i$ and $\bf{S}_i$ are the maximum likelihood estimates of $\bs\mu_i$ and $\bs\Sigma_i$, respectively. For this training-data scenario, the supervised \textit{QDA} decision rule uses \eqref{eq:d_hat} rather than  $d_{i}(\bf{x})$ in \eqref{eqn:QDA}. 

\subsection{An SDR Theorem}\label{subsec:SDR_model}
\noindent We now present a theoretical result that yields an \textit{SDR}-based \textit{LDR} transformation of the full-dimensional feature data that preserves the optimal error rate associated with the full-dimensional feature-space under the \textit{QDA} decision rule when all population mean vectors and covariance matrices are known.  This result is motivated by the concept from \cite{peters1978} of linear sufficiency for differences in $k$ heteroscedastic multivariate populations.   \cite{ounpraseuth2015} have   derived similar results under the multivariate normality assumption. However, our results, presented in the theorem below and the lemmas in the appendix, are distribution-free. Additionally, we make no assumption concerning prior probabilities for classes.  We remark that  \cite{xie2007} have found that unequal class prior probabilities negatively affect the performance of  \textit{QDA} in terms of the \textit{Area Under the Receiver Operating Characteristic Curve} (\textit{AUC}).  However,  \cite{xue2008} have demonstrated that although re-balancing training-sample sizes may lead to a marginal increase in the \textit{AUC}, a relatively large increase in the \textit{CER} often follows.

The \textit{SDR} approach presented in the theorem provides necessary and sufficient conditions such that classification of an unlabeled observation into a population remains invariant when we use the linearly-transformed lower-dimensional data.

\begin{thm*}\label{Thm1}
Suppose we have $k$ multivariate populations with full classification information contained in the population means $\bs\mu_i$ and population covariance matrices $\bs\Sigma_i \in \mathbb{S}^{p}_{+}$, where $i = 1, \ldots, k$.  Let $\pi_i$ denote \emph{a priori} class membership, and let
\begin{equation} \label{eq:M}
\bf{M} \coloneqq \left[\bs\Sigma^{-1}_2\bs\mu_2 - \bs\Sigma^{-1}_1\bs\mu_1 |\ldots| \bs\Sigma^{-1}_k\bs\mu_k - \bs\Sigma^{-1}_1\bs\mu_1 | \bs\Sigma_2 - \bs\Sigma_1|\ldots| \bs\Sigma_k - \bs\Sigma_1\right],
\end{equation}
where $\bf{M} \in \R^{p \times s}$, $s = (k - 1)(p + 1)$ with $\text{rank}(\bf{M}) = q < p$. Let $\bf{M} =\bf{U} \bf{D}\bf{V}^{T}$  $\in\R^{p \times s}$ be the \textit{reduced SVD} of $\bf{M}$, where $\bf{U} \in \R^{p \times q}$, $\text{rank}(\bf{U}) = q < p$, $\bf{D} \in R^{q \times q}$, and $\bf{V} \in R^{s \times q}$.  Then, for an unlabeled observation vector $\bf{x} \in \R^{p}$, $D(\bf{x}) = D\left(\bf{U}^{T}\bf{x}\right)$, where $D(\bf{x})$ is defined in \eqref{eqn:QDA}.

\begin{proof}
First let $\bf{U} \in \R^{p \times q}$ with rank$(\bf{U}) = q$. Now let $\bf{P}_\bf{U}^{\perp} \coloneqq \left(\bf{I} - \bf{UU}^{T}\right)$ and $\bf{C} \coloneqq \bf{R}\bf{P}_\bf{U}^{\perp}$,  where $\bf{R}\in\R^{(p - q) \times p}$ and rank$(\bf{R}) = p - q$. Using Lemmas \ref{lemma1}, \ref{lemma2}, and \ref{lemma3} in the appendix and letting $\bf{A} = \left[\bf{U}^{T},\bf{C}^T \right]^T \in \mathbb{S}^{p}_{+}$, we have that 

\begin{align*}
d_i(\bf{x}) &= d_i(\bf{Ax}) \\
&=- 2\log\left( \pi_i \right) + \log\left|\bf{A}\bs\Sigma\bf{A}^T\right| +  \left[\bf{A}\left(\bf{x} - \bs\mu_i\right)\right]^T \left(\bf{A}\bs\Sigma_i \bf{A}^T\right)^{-1} \left[\bf{A}\left(\bf{x} -\bs\mu_i\right)\right] \\
&= - 2 \log\left(\pi_i \right) + 
\log\begin{vmatrix} 
\bf{U}^{T}\bs\Sigma_i\bf{U} & \bf{U}^{T}\bs\Sigma_i \bf{C}^T  \\ \bf{C}\bs\Sigma_i\bf{U}  &  \bf{C}\bs\Sigma_i \bf{C}^T 
\end{vmatrix} \\
&\qquad + \begin{bmatrix} \bf{U}^{T}\left(\bf{x} - \bs\mu_i \right) \\ \bf{C}\left(\bf{x} - \bs\mu_i \right) \end{bmatrix}^T \begin{bmatrix} \bf{U}^{T}\bs\Sigma_i\bf{U} & \bf{U}^{T}\bs\Sigma_i\bf{C}^T  \\  \bf{C}\bs\Sigma_i\bf{U}  &  \bf{C} \bs\Sigma_i \bf{C}^T \end{bmatrix}^{-1} \begin{bmatrix} \bf{U}^{T}\left(\bf{x} - \bs\mu_i \right) \\ \bf{C}\left(\bf{x} - \bs\mu_i \right)  \end{bmatrix} \\
&= - 2 \log\left(\pi_i\right) + 
    \log\begin{vmatrix} 
        \bf{U}^{T}\bs\Sigma_i\bf{U} & \bf{0}  \\ 
        \bf{0} & \bf{C} \bs\Sigma_i \bf{C}^T
    \end{vmatrix} \\
&\qquad +
    \begin{bmatrix} 
        \bf{U}^{T}\left(\bf{x} - \bs\mu_i \right) \\ 
        \bf{C}\left(\bf{x} - \bs\mu_i \right) \end{bmatrix}^T 
    \begin{bmatrix} 
        \bf{U}^{T}\bs\Sigma_i\bf{U} & \bf{0} \\ 
        \bf{0} & \bf{C} \bs\Sigma_i \bf{C}^T 
    \end{bmatrix}^{-1} 
    \begin{bmatrix} 
        \bf{U}^{T}\left(\bf{x} - \bs\mu_i \right) \\ 
        \bf{C}\left(\bf{x} - \bs\mu_i \right) 
    \end{bmatrix}  \\
&= - 2\log\left(\pi_i\right) + \log\left| \bf{U}^{T}\bs\Sigma_i\bf{U} \right| + \log\left| \bf{C} \bs\Sigma_1 \bf{C}^T  \right| \\
&\qquad + \left[\bf{U}^{T} \left(\bf{x} - \bs\mu_i\right)\right]^T \bf{U}^T \bs\Sigma^{-1}_i \bf{U} \left[\bf{U}^{T}\left(\bf{x} - \bs\mu_i \right)\right]\\
&\qquad + \left[\bf{C} \left(\bf{x} - \bs\mu_1 \right)\right]^T \left[\bf{C} \bs\Sigma_1 \bf{C}^T \right]^{-1} \left[\bf{C} \left(\bf{x} - \bs\mu_1 \right)\right]  \\
&= - 2\log\left(\pi_i\right) + \log\left| \bf{U}^{T}\bs\Sigma_i\bf{U} \right| + \left[\bf{U}^{T} \left(\bf{x} - \bs\mu_i\right)\right]^T \bf{U}^T \bs\Sigma^{-1}_i \bf{U} \left[\bf{U}^{T} \left(\bf{x} - \bs\mu_i\right)\right] + c \\
&= d_i\left(\bf{U}^{T}\bf{x}\right) + c,
\end{align*}
where $c \coloneqq \log\left| \bf{C} \bs\Sigma_1 \bf{C}^T \right| + \left[\bf{C} \left(\bf{x} - \bs\mu_1 \right)\right]^T \left[\bf{C} \bs\Sigma_1 \bf{C}^T \right]^{-1} \left[\bf{C} \left(\bf{x} - \bs\mu_1 \right)\right]$. Because $c$ is fixed for the reference parameters $\bs\mu_1$ and $\bs\Sigma_1$, we have  $d_i(\bf{x}) > d_j(\bf{x})$ if and only if $d_i\left(\bf{U}^{T}\bf{x}\right) > d_j\left(\bf{U}^{T}\bf{x}\right)$ for all $i,j = 1, \ldots, k$; $i \ne j$. Therefore, $D(\bf{x}) = D\left(\bf{U}^{T}\bf{x}\right)$.  The arguments are reversible and, thus, the theorem holds.
\end{proof}
\end{thm*}

In the above theorem, assuming all parameters are known, the \textit{DRS} is $\mathcal{S}_{\bf{M}} \coloneqq \text{span}(\bf{M})$. Because $\bf{U}^{T}$ provides a basis for $\mathcal{S}_{\bf{M}} \subseteq \mathcal{S}_{Y|\bf{X}}$, a relationship guaranteed by the linearity condition of $E[\bf{X}|\bf{U}^{T}\bf{X}]$, we have derived an \textit{SDR} projection matrix that preserves the full-dimensional optimal error rate under the \textit{QDA} decision rule. Thus, we can replace an unlabeled vector $\bf{x} \in \R^{p}$ with the lower-dimensional  vector $\bf{U}^{T}\bf{x} \in \R^{q}$, $q < p$, with no loss of supervised classification information.

\subsection{An SDR Method with Precision Matrix Shrinkage Estimators}\label{subsec:SDR_Shrinkage}
\noindent In Section \ref{sec:intro}, we discuss the benefits of \textit{SDR} methods in terms of efficiency for parameter estimation. Moreover, we seek to improve upon the use of \textit{SDR} in supervised classification by avoiding the over-parametrization that often occurs with a relatively small training-sample size. We can achieve this improvement by reducing the variability of the class-specific precision matrix estimators for $\bs{\Omega}_{i} = \bs{\Sigma}^{-1}_{i}$. Thus, motivated by the theorem in Section \ref{Thm1}, we propose a new \textit{SDR} method that adds stability to the \textit{LDR} projection matrix through induced sparsity and bias by utilizing a specified precision shrinkage estimator. Let $\widehat{\bs{\Omega}}_{i}$ be an estimator of $\bs{\Omega}_{i}$. Then when the parameters are unknown, we replace the columns of $\bf{M}$ in \eqref{eq:M} with their sample estimates and user-specified precision estimators, yielding
\begin{equation} \label{eq:Mhat}
\widehat{\bf{M}} \coloneqq \left[\widehat{\bs{\Omega}}_2\bar{\bf{x}}_2 - \widehat{\bs{\Omega}}_1\bar{\bf{x}}_1|\ldots|\widehat{\bs{\Omega}}_k\bar{\bf{x}}_k - \widehat{\bs{\Omega}}_1\bar{\bf{x}}_1|\bf{S}_2 -  \bf{S}_1|\ldots|\bf{S}_k - \bf{S}_1\right].
\end{equation}
In the theorem, we show that multiplying an unlabeled observation $\bf{x} \in \R^{p}$ by the \textit{SDR}-based \textit{LDR} projection matrix $\bf{U}^{T} \in \R^{q \times p}$ preserves all classification information in a reduced $q$-dimensional subspace. However, we cannot directly obtain $\bf{U}^T$ when rank$(\widehat{\bf{M}}) = p$. Moreover, we may wish to obtain a lower dimensional representation of the original data with dimension $r$, where $1 \leq r \leq q < p$.  Thus, we look to construct an $r$-dimensional \textit{SDR}-based \textit{LDR} projection matrix that preserves all of the original $p$-dimensional classification information. Let 
$\bf{U}_{(r)} \in \R^{p \times r}$ denote the matrix composed of the $r$ vectors of $\bf{U}$ corresponding to the $r$ largest singular values in the \textit{SVD}. Finally, let $\widehat{\bf{M}} = \bf{U}_{\widehat{\bf{M}}}\bf{D}_{\widehat{\bf{M}}}\bf{V}^{T}_{\widehat{\bf{M}}}$ be the \textit{SVD} of \eqref{eq:Mhat}.  Then we define $\bf{U}^{T}_{\widehat{\bf{M}}, (r)} \in \R^{r \times p}$ to be the $r$-dimensional \textit{SDR}-based \textit{LDR} projection matrix for reducing the feature dimension from dimension $p$ to the smaller dimension $r$.  We refer to this \textit{SDR} method as \textit{stabilized SDR} (\textit{SSDR}).

For supervised classification purposes, 
we project an unlabeled observation $\bf{x} \in \R^{p}$ by applying the linear transformation $\bf{U}^{T}_{\widehat{\bf{M}},(r)}{\bf{x}} \in \R^{r}$ and then applying a specified classification rule to the resulting r-dimensional reduced data. Our \textit{SSDR} approach provides a general \textit{SDR} technique usable with any supervised classification method. However, here, we adopt the supervised \textit{QDA} decision rule in \eqref{eq:d_hat} for classification because our \textit{SSDR} method is motivated by the theorem in Section \ref{subsec:SDR_model} and incorporates heteroscedastic information. Lastly, for selecting the optimal dimension $r$, we propose the estimated \textit{CER} using a cross-validation method as the criterion. 

\section{Four Precision Matrix Shrinkage Estimators}\label{sec:estimators}

\noindent In this section, we describe four precision matrix shrinkage estimators proposed by \cite{molstad2018}, \cite{haff1979}, \cite{wang2015},  and \cite{bodnar2016} that we use to formulate our proposed \textit{SSDR} method, described in Section \ref{subsec:SDR_Shrinkage}. We refer to these four precision matrix estimators as the \textit{MRY}, \textit{Haff}, \textit{Wang}, and \textit{Bodnar} estimators, respectively. 

\subsection{MRY Estimator}\label{sub:molstad}
\noindent The first precision matrix shrinkage estimator we consider was proposed by \cite{molstad2018} and is given by 
\begin{equation}\label{eq:Molstad}
    \widehat{\bs{\Omega}}_{i} \coloneqq \underset{\bs{\Omega}_{\ast} \in \mathbb{S}^{p}_{+}}{\text{arg min}} \left\{ tr(\bf{S}_{i}\bs{\Omega}_{\ast}) - \log\left|\bs{\Omega}_{\ast}\right| + \lambda^{\ast}_{i} |\bf{A}_{i}\bs{\Omega}_{\ast} \bf{B}_{i} - \bf{C}_{i}|_{1}\right\}, 
\end{equation}
where $\lambda^{\ast}_{i} > 0$ is a tuning parameter, and $\bf{A}_{i} \in \R^{a \times p}$, $\bf{B}_{i} \in \R^{p \times b}$, and $\bf{C}_{i} \in \R^{a \times b}$ are user-specified matrices. Also, $|M|_{1} = \sum_{l \neq j}|M_{lj}|$ is an $L_{1}$-like norm that forces sparsity on the off-diagonal elements. This shrinkage estimator exploits the assumption that $\bf{A}_{i} \bs\Omega_{i} \bf{B}_{i} - \bf{C}_{i}$ is sparse and uses $\bf{C}_{i}$  as a shrinkage target matrix for $\bf{A}_{i} \bs\Omega_{i} \bf{B}_{i} - \bf{C}_{i}$.

For \textit{QDA}, \cite{molstad2018} have proposed the specified matrices $\bf{C}_{i} = \bs{0}_{p}$, a $p \times p$ matrix of zeros, and $\bf{A}^{\text{T}}_{i} = \bf{B}_{i} = (\bar{\bf{x}}_{i}, \gamma_{i} \bf{I}_{p})$ for the tuning parameter $\gamma_{i} > 0$, assuming $\bs{\mu}^{\text{T}}_{i}\bs{\Omega}_{i}\bs{\mu}_{i}$ is small. That is, $\bs{\mu}_{i}$ is in the span of a set of the eigenvectors corresponding to the smallest eigenvalues of $\bs{\Omega}_{i}$. However, selecting the tuning parameters using the user-specified matrices for \textit{QDA} recommended by \cite{molstad2018} becomes computationally cumbersome as one increases the number of classes. Thus, for simplicity we also consider the specified matrices to be $\bf{C}_{i} = \bs{0}_{p}$ and $\bf{A}_{i} = \bf{B}_{i} = \bf{I}_{p}$,
which becomes the popular lasso precision matrix  shrinkage estimator 
\begin{equation}\label{eq:yuan}
    \widehat{\bs{\Omega}}_{i} \coloneqq \underset{\bs{\Omega}_{\ast} \in \mathbb{S}^{p}_{+}}{\text{arg min}} \left\{ tr(\bf{S}_{i}\bs{\Omega}_{\ast}) - \log\left|\bs{\Omega}_{\ast}\right| + \lambda^{\ast}_{i} |\bs{\Omega}_{\ast}|_{1}\right\} 
\end{equation}
proposed by \cite{yuan2007}.  To solve the optimization in \eqref{eq:Molstad} and \eqref{eq:yuan}, \cite{molstad2018} proposed an alternating direction method-of-multipliers algorithm based on the majorize-minimize principle of \cite{lange2016}.

\subsection{Haff Estimator}
\noindent We next consider the precision matrix shrinkage estimator  proposed by \cite{haff1979}, which induces bias to stabilize the estimator of $\bs{\Omega}_{i}$ and is given by 

\begin{equation}\label{eq:haff}
    \widehat{\bs{\Omega}}_{i} \coloneqq (1 - t(U_i))(n_i - p - 2)\bf{S}_i^{-1} + \frac{t(U_i)(pn_i - p - 2)}{tr(\bf{S}_i)}\bf{I}_{p}, 
\end{equation}
where 
\[
    t(U_i) \coloneqq \text{min} \left \{ \frac{4(p^2 - 1)}{(n_i - p -2)p^2}, 1 \right \} U_i^{1/p}
\]
and 
\[
U_i \coloneqq \frac{p|\bf{S}_i|^{1/p}}{tr(\bf{S}_i)}.
\]
Here, the ratio $U_{i}$ quantifies the disparity among the eigenvalues of $\bf{S}_{i}$, with the numerator and denominator representing the geometric and arithmetic means, respectively. The function $t(U_{i})$ serves as the shrinkage factor, controlling the degree of shrinkage applied to $\bf{S}_{i}$.

\subsection{Wang Estimator}
\noindent The precision matrix  estimator proposed by \cite{wang2015} is the  ridge-type shrinkage estimator given by
\begin{equation}\label{eq:wang}
    \widehat{\bs{\Omega}}_{i} \coloneqq \widehat{\alpha}_{i}(\bf{S}_{i} + \widehat{\beta}_{i}\bf{I}_{p})^{-1}, 
\end{equation}
where $\hat{\alpha}_{i} > 0$ and $\hat{\beta}_{i} > 0$ are shrinkage coefficients under the loss function 
\begin{equation}\label{eq:loss}
\frac{1}{p}tr(\widehat{\bs{\Omega}}_{i}\bs{\Sigma}_{i} - \bf{I}_{p})^2.
\end{equation}
This loss function has been utilized by \cite{haff2}, \cite{k_and_g}, and  \cite{yang1994}.  We estimate \eqref{eq:loss} with the empirical loss function
\begin{equation}\label{eq:emp_loss}
   L_{i}(\beta) \coloneqq 1 - \frac{(\widehat{R}_{1}(\beta))^{2}}{\widehat{R}_{2}(\beta)},  
\end{equation}
where
\begin{align*}
    \widehat{R}_{1}(\beta) &\coloneqq \frac{a_{1}(\beta)}{1 - (p/n_{i})a_{1}(\beta)},\\
    \widehat{R}_{2}(\beta) &\coloneqq \frac{a_{1}(\beta)}{(1 - (p/n_{i})a_{1}(\beta))^{3}} - \frac{a_{2}(\beta)}{(1 - (p/n_{i})a_{1}(\beta))^{4}},
\end{align*}
and $a_{1}(\beta) \coloneqq 1 - (1/p)tr((1/\beta)\bf{S}_{i} + \bf{I}_{p})^{-1}$ and $a_{2}(\beta) \coloneqq (1/p)tr((1/\beta)\bf{S}_{i} + \bf{I}_{p})^{-1} - (1/p)tr((1/\beta)\bf{S}_{i} + \bf{I}_{p})^{-2}$. By minimizing the empirical loss in \eqref{eq:emp_loss}, we determine that 
\[
\widehat{\beta}_{i} \coloneqq \underset{\beta_{i} \in [\lambda_{min}, \lambda_{max}]}{\text{arg min}}L_{i}(\beta),
\]
and
\[
\widehat{\alpha}_{i} \coloneqq \frac{\widehat{R}_{1}(\widehat{\beta}_{i})}{\widehat{R}_{2}(\widehat{\beta}_{i})},
\]
where $\lambda_{min}$ and $\lambda_{max}$ are the extrema eigenvalues for $\bf{S}_{i}$. In the case that $\hat{\beta}_{i}$ is not unique, we take $\hat{\beta}_{i}$ to be the smallest solution.

\subsection{Bodnar Estimator}
\noindent \cite{bodnar2016} have proposed a  precision matrix estimator that shrinks toward some user-specified symmetric positive-definite target matrix $\bs{\Omega}_0$. Let $||\bf{A}||^{2}_{\text{F}} = tr(\bf{A}\bf{A}^{\prime})$ denote the Frobenius norm of a square matrix $\bf{A}^{p \times p}$, and let $||\bf{A}||_{\text{tr}} = tr((\bf{A}\bf{A}^{\prime})^{1/2})$ be the trace norm. Then, for the case where the parameters are unknown, the precision matrix estimator proposed by \cite{bodnar2016} is
\begin{equation}\label{eq:bodnar}
    \widehat{\bs{\Omega}}_{i} \coloneqq \widetilde{\alpha}_{i}\bf{S}^{-1}_{i} + \widetilde{\beta}\bs{\Omega}_0, \text{ for } \underset{p}{\text{sup}}\frac{1}{p}||\bs{\Omega}_{0}||_{\text{tr}} \leq M, 
\end{equation}
where 
\[
\widetilde{\alpha}_i \coloneqq 1 - p/n - \frac{(1/n)||\bf{S}^{-1}_{i}||^{2}_{\text{tr}}||\bs{\Omega}_{0}||^{2}_{\text{F}}}{||\bf{S}^{-1}_{i}||^{2}_{\text{F}}||\bs{\Omega}_{0}||^{2}_{\text{F}} - (tr(\bf{S}^{-1}_{i}\bs{\Omega}_{0}))^{2}},
\]
and
\[
\widetilde{\beta}_i \coloneqq \frac{tr(\bf{S}^{-1}_{i}\bs{\Omega}_{0})}{||\bs{\Omega}_{0}||^{2}_{\text{F}}}(1 - p/n - \widetilde{\alpha}_{i}).
\]

For the target matrix, we take $\bs{\Omega}_{0} = \bf{I}_{p}$, which \cite{bodnar2016} proposed as a naive prior when no information is available concerning $\bs{\Sigma}_{i}$. We can utilize information on the precision structure in $\bs{\Omega}_{0}$, such as when using $diag(\bf{S}_{i})$, to yield sparsity. To ensure the coefficient $\hat{\beta}_{i}$ is bounded for large dimensions $p$, we need only assume that the target matrix $\bs{\Omega}_{0}$ is a uniformly-bounded trace norm.  That is, an $M > 0$  exists such that $\text{sup}_{p}(1/p)||\bs{\Omega}_{0}||_{\text{tr}} \leq M$.

\section{A Monte-Carlo Simulation Contrast of Precision Matrix Estimators for SDR} \label{sec:mc_sim}

\subsection{Monte Carlo Simulation Design}\label{subsec:mc_sim_descr}
\noindent Here, we describe the \textit{MC} simulation design that we used to contrast the classification efficacy of the precision matrix estimators in \eqref{eq:Molstad}, \eqref{eq:haff}, \eqref{eq:wang},  and \eqref{eq:bodnar} in Section \ref{sec:estimators} in our proposed \textit{SSDR} approach  described in Section \ref{subsec:SDR_Shrinkage}. For a baseline, we also considered the sample precision matrix estimator, $\bf{S}^{-1}_{i}$. We refer to these \textit{SSDR}-based  methods as $SSDR_{MRY}$, $SSDR_{Haff}$, $SSDR_{Bod}$, $SSDR_{Wang}$, and  $SSDR_{S^{-1}}$. For each \textit{SSDR} method, we use \textit{QDA} as the supervised classifier. 

We considered four parameter configurations from multivariate normal populations for our MC simulation study.  These are  similar to the parameter configurations found in \cite{ounpraseuth2015} and \cite{wu2022} and are given below.

\begin{itemize}
    \item \textit{Configuration 1}: Different mean vectors with identical spherical covariance matrices and $p = 10$, $\bs{\Sigma}_{1} = \bs\Sigma_{2} = \bf{I}_{p}$ and $\bs\mu_{1} = \bs{0}_{p},\ \bs\mu_{2} = \bs{1}_{p}$.
    \item \textit{Configuration 2}: Different covariance matrices consisting of spherical, intra-class, and spiked covariance matrix  structures with $p = 10$. $\bs{\Sigma}_{1} = \bf{I}_{p},\ \bs{\Sigma}_{2} = \bs{\Sigma}_{1} + \bf{J}_{p}, $ where $\bf{J}_{p}$ is a $p \times p$ matrix of ones,

\[
    \bs{\Sigma}_{3} = \begin{bmatrix}
            2 & 1 & 0 & 1 & 1 & 1 & 1 & 1 & 1 & 1  \\
            1 & 2 & 0 & 1 & 1 & 1 & 1 & 1 & 1 & 1  \\
            0 & 0 & 10 & 0 & 0 & 0 & 0 & 0 & 0 & 0 \\
            1 & 1 & 0 & 2 & 1 & 1 & 1 & 1 & 1 & 1   \\
            1 & 1 & 0 & 1 & 2 & 1 & 1 & 1 & 1 & 1   \\
            1 & 1 & 0 & 1 & 1 & 2 & 1 & 1 & 1 & 1  \\
            1 & 1 & 0 & 1 & 1 & 1 & 2 & 1 & 1 & 1 \\
            1 & 1 & 0 & 1 & 1 & 1 & 1 & 2 & 1 & 1 \\
            1 & 1 & 0 & 1 & 1 & 1 & 1 & 1 & 2 & 1 \\
            1 & 1 & 0 & 1 & 1 & 1 & 1 & 1 & 1 & 2 \\
    \end{bmatrix},
\]
\text{and}    $\bs\mu_{1} = (-1.43, -0.66, -0.94, 0.31, -0.19, 0.89, 0.25, -0.34, 1.25, -1.60)^{\prime},\ \bs\mu_{2} = \bs{\mu}_{1} + \bs{1}_{p},\ \text{and} \ \bs\mu_{3} = \bs{\mu}_{1} + \bs{2}_{p}.$
    \item \textit{Configuration 3}: Two identical covariance matrices with $p = 10$, 
    \[
    \bs{\Sigma}_{1} = \bs{\Sigma}_{2} = \begin{bmatrix}
                10 & 4 & 5 & 4 & 3 & 4 & 4 & 5 & 4 & 3 \\
                4 & 10 & 5 & 2 & 4 & 3 & 3 & 5 & 4 & 3 \\
                5 & 5 & 10 & 5 & 5 & 3 & 4 & 4 & 4 & 4 \\
                4 & 2 & 5 & 10 & 3 & 4 & 2 & 3 & 4 & 3 \\
                3 & 4 & 5 & 3 & 12 & 3 & 4 & 5 & 3 & 3 \\
                4 & 3 & 3 & 4 & 3 & 9 & 3 & 4 & 4 & 4 \\
                4 & 3 & 4 & 2 & 4 & 3 & 14 & 2 & 2 & 2 \\
                5 & 5 & 4 & 3 & 5 & 4 & 2 & 12 & 1 & -0.5 \\
                4 & 4 & 4 & 4 & 3 & 4 & 2 & 1 & 14 & -1 \\
                3 & 3 & 4 & 3 & 3 & 4 & 2 & -0.5 & -1 & 11
                \end{bmatrix},
    \]
\text{and} $\bs{\Sigma}_{3} = \bf{I}_{p}$, with $\bs\mu_{1} = \bs{0}_{p}$ $\bs\mu_{2} = \bs{5}_{p}$ and $\bs\mu_{3} = \bs{10}_{p}.$
    \item \textit{Configuration 4}: Covariance matrices consisting of spherical and intra-class structures except that $p = 50$. $\bs{\Sigma}_{1} = \bf{I}_{p}$, $\bs{\Sigma}_{2} = \bs{\Sigma}_{1} + 2\bf{J}_{p}$, $\bs\mu_{1} = \bs{0}_{p}$, and $\bs\mu_{2}$ is a $p \times 1$ vector with \textit{IID} entries from the uniform(0, 1) distribution.
\end{itemize}

\begin{figure}[t!]
    \centering
    \includegraphics{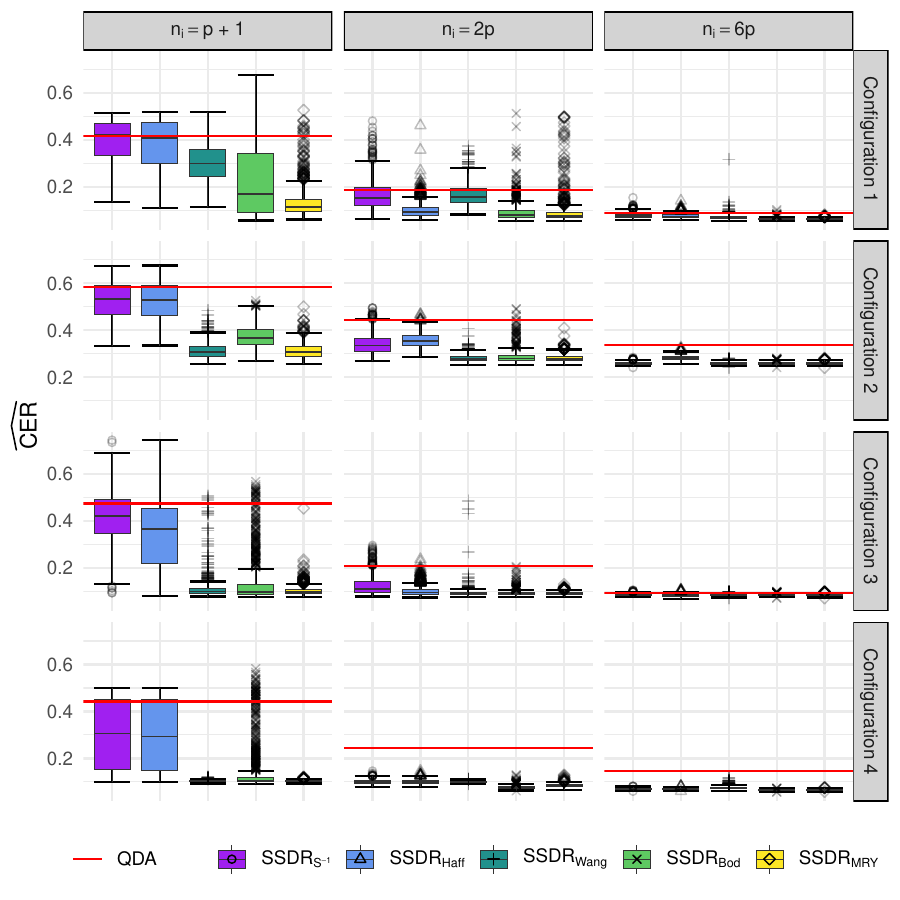}
    \caption{\textit{Estimated conditional error rate}, denoted $\widehat{CER}$, plots for the simulation described in Section \ref{subsec:mc_sim_descr} for contrasting the \textit{SSDR} methods. We display the $\widehat{CER}$s for the dimension that achieved the global minimum median $\widehat{CER}$ for each  method, parameter configuration, and class-specific training-sample size. The full-feature median $\widehat{CER}$ using \textit{QDA} is represented by the horizontal line.}
    \label{fig:1}
\end{figure}

\noindent In Configurations 2 and 4, for the ${SSDR_{MRY}}$ method, we adopted the user-specified matrices proposed for  \textit{QDA} by \cite{molstad2018} given in Section \ref{sub:molstad}. For Configurations 1 and 3, the \textit{QDA}-recommended matrices performed the same or worse in terms of the estimated \textit{CER} when contrasted to the simple specified matrices of $\bf{C}_{i} = \bs{0}_{p}$ and $\bf{A}_{i} = \bf{B}_{i} = \bf{I}_{p}$. For Configurations 1 and 3, we used the simple user-specified matrices, which simplified \eqref{eq:Molstad} to \eqref{eq:yuan}. For the $SSDR_{MRY}$  approach, we determined the tuning parameters by minimizing the validation \textit{CER}s across a grid search informed by the eigenvalues of the class-specific sample precision matrices. For each configuration, we generated 5,000 observations from each multivariate normal distribution. We then varied the class-specific training-sample sizes  $n_{i} = p + 1,\ 2p,\ 6p, i = 1,\ldots,k,$ to simulate poorly-posed to well-conditioned covariance matrices. We held the remaining observations per class aside as a test set. For each \textit{SSDR} method, we  projected the training and test sets from $p$ to $r$ dimensions and then used \textit{QDA} as the classification rule and recorded the \textit{estimated CER}, denoted $\widehat{CER}$. We replicated this process 1,000 times for each configuration. For Configuration 4, we generated the parameters once and independently drew all data replicates from the same distribution. Because of the skewness of the distribution of the $\widehat{CER}$s, we evaluated the classification efficacy in terms of the median of the $\widehat{CER}$s, denoted as $\widetilde{CER}$. We represented the variability in the $\widehat{CER}$s by the \textit{standard deviation} (\textit{SD}). Also, we used $\widetilde{CER}_{F}$ to denote the median of the $\widehat{CER}$s of the full-feature data using only \textit{QDA} and no dimension reduction. We have summarized our $MC$ simulation results in Figure \ref{fig:1} by displaying the distribution of the $\widehat{CER}$s for the \textit{DRS} that achieved the global minimum $\widetilde{CER}$ for each  class-specific training-sample size per parameter configuration. Additionally, the $\widetilde{CER}_{F}$ is given by the horizontal line. 

\subsection{Monte Carlo Simulation Results}

\noindent For Configuration 1, we recall that $p = 10$ and the optimal reduced-subspace dimension was $q = 1$. For the full-feature \textit{QDA}, we determined that the $\widetilde{CER}_{F}s$ for the cases $n_{i} = p + 1,\ 2p,\ \text{and}\  6p$ were 0.4167, 0.1858, and 0.0866, respectively. For the case $n_{i} = p+1 = 11$, the full-data sample covariance matrices were poorly posed because an insufficient number of sample observations were used to inadequately estimate the number of free parameters. Consequently, using the $SSDR_{S^{-1}}$ method resulted in approximately the same $\widetilde{CER}$ as $\widetilde{CER}_{F}$. Moreover, we found that $SSDR_{Haff}$ performed similarly to the $SSDR_{S^{-1}}$ and provided no significant reduction of $\widetilde{CER}_{F}$. The $SSDR_{Wang}$, $SSDR_{Bod}$, and $SSDR_{MRY}$ methods achieved their respective global minimum $\widetilde{CER}$s of 0.2990 at $r = 6$, 0.1696 at $r = 1$, and 0.1138 at $r = 2$, respectively, per method. Thus, for the given training-sample size, the $SSDR_{MRY}$  method  significantly reduced the $\widetilde{CER}_{F}$ by 0.3029.  Although the $SSDR_{Bod}$   notably reduced the $\widetilde{CER}_{F}$ for this training-sample size, its \textit{CER} exhibited large variability with a \textit{SD} of 0.1551. For $n_{i} = 2p = 20$,  only the $SSDR_{Wang}$ , $SSDR_{Bod}$ , and $SSDR_{MRY}$  yielded notable reductions to the $\widetilde{CER}_{F}$ with $\widetilde{CER}$s of 0.0914 at $r = 7$, 0.0787 at $r = 1$, and 0.0768 at $r = 1$, respectively. In particular, the $SSDR_{Bod}$ and $SSDR_{MRY}$ methods  reduced $\widetilde{CER}_{F}$ by   0.1000. For the case when $n_{i} = 6p = 60 > p^{2}/2$, we usually would not have expected smaller $\widetilde{CER}$s when  contrasted to the $\widetilde{CER}_{F}s$ for the \textit{SDR} methods considered here. However, the $SSDR_{Bod}$  and $SSDR_{MRY}$  approaches still  reduced the $\widetilde{CER}_{F}$ by 0.0618 and 0.0615, respectively, for the reduced dimension $r = 1$. In addition, they yielded small \textit{SD}s of 0.0047 and 0.0037, respectively. 

For Configuration 2, we had $p = 10$ and $q = 2$.  Using the full-feature \textit{QDA}, we obtained the  $\widetilde{CER}_{F}$s corresponding to the class-specific training-sample sizes $n_{i} = p + 1,\ 2p,\ \text{and}\ 6p$ of 0.5856, 0.4433, and 0.3379, respectively. For the case where $n_{i} = p + 1$, the $SSDR_{S^{-1}}$ and $SSDR_{Haff}$ methods showed practically insignificant reductions to  $\widetilde{CER}_{F}$. However, the $SSDR_{Wang}$, $SSDR_{Bod}$, and $SSDR_{MRY}$  methods resulted in minimum $\widetilde{CER}$s of 0.3078 at $r = 2$, 0.3688 at $r = 3$, and 0.3083 at $r = 2$, respectively. Thus, for this training-sample size scenario, the $SSDR_{Wang}$ and $SSDR_{MRY}$ methods yielded the largest reduction of $\widetilde{CER}_{F}$, approximately 0.2770. Also, the $SSDR_{Wang}$ method had the smallest \textit{SD} of 0.0311. For the case where $n_i = 2p$, the $SSDR_{Wang}$, $SSDR_{Bod}$, and $SSDR_{MRY}$ methods provided approximately the same $\widetilde{CER}$ of approximately 0.2800 at $r = 2$, resulting in a reduction to $\widetilde{CER}_{F}$ of approximately 0.1633. Similarly, for the case where $n_i = 6p$, the $SSDR_{S^{-1}}$, $SSDR_{Wang}$, $SSDR_{Bod}$, and $SSDR_{MRY}$ methods yielded approximately the same $\widetilde{CER}$ of about 0.2590, all at $r = 2$, resulting in a reduction to $\widetilde{CER}_{F}$ of approximately 0.0789. For the two cases when $n_i = 2p$ and $6p$, the $SSDR_{Haff}$ method yielded a slightly increased $\widetilde{CER}$ when contrasted to $SSDR_{S^{-1}}$.

For Configuration 3, $p = q = 10$. For $n_{i} = p + 1,\ 2p,\ \text{and}\ 6p$, the $\widetilde{CER}_{F}$s were 0.4743, 0.2065, and 0.0917. We remark that although $\bf{M}$ was full rank, each of our proposed \textit{SSDR}-based  methods achieved smaller $\widetilde{CER}$s than $\widetilde{CER}_{F}$ for all $r < 10$. Specifically, for the case when $n_{i} = p + 1$, all five \textit{SSDR}-based methods achieved their minimum $\widetilde{CER}$ at $r = 1$. For this training-sample size, the $SSDR_{Wang}$, $SSDR_{Bod}$, and $SSDR_{MRY}$ methods achieved respective minimum $\widetilde{CER}$s of 0.0996, 0.0958, and 0.0962. Thus, these three \textit{SSDR} methods yielded a notable reduction of approximately 0.3700 to $\widetilde{CER}_{F}$. However, the $SSDR_{Bod}$ technique resulted in a \textit{SD} of 0.1017, which was greater than its $\widetilde{CER}$. In contrast, the \textit{SD} for the $SSDR_{MRY}$ was 0.0217. Again, we observed that the $SSDR_{S^{-1}}$ and $SSDR_{Haff}$ methods performed similarly to each other with $\widetilde{CER}$s that yielded little reduction to the $\widetilde{CER}_{F}$. For $n_{i} = 2p$ and $6p$, all \textit{SSDR}  methods yielded similar $\widetilde{CER}$s between 0.0800 and 0.0900. The two smallest $\widetilde{CER}$s per training-sample size were 0.0892 by $SSDR_{MRY}$ at $ r = 1$ and 0.0804 by $SSDR_{Haff}$ at $r = 4$. 

In Configuration 4, we had $p = 50$ and $q = 2$. For $n_{i} = p + 1,\ 2p,\ \text{and}\ 6p$, the $\widetilde{CER}_{F}$s were 0.4426, 0.2458, and 0.1449, respectively. For $n_{i} = p + 1 = 51$, the $SSDR_{S^{-1}}$ and $SSDR_{Haff}$ methods yielded similar $\widetilde{CER}s$ of approximately 0.3000, both at $r = 2$. Thus, these $\widetilde{CER}s$ reduced the $\widetilde{CER}_{F}$ by about 0.1426. However, both of these methods exhibited high variability in \textit{CER}s with \textit{SD}s of approximately 0.1443. The $SSDR_{Wang}$ and $SSDR_{MRY}$ methods achieved the same minimum $\widetilde{CER}$ of 0.1030 at $r = 1$, thus yielding a reduction in $\widetilde{CER}_{F}$ of 0.3396. Moreover, each of these two \textit{SSDR} approaches yielded small \textit{SD}s of 0.0041. The $SSDR_{Bod}$ method achieved a similar $\widetilde{CER}$ as the $SSDR_{Wang}$  and $SSDR_{MRY}$ but again exhibited a considerably larger \textit{SD} of 0.1034. For $n_{i} = 2p = 100$, all considered \textit{SSDR} methods notably reduced the $\widetilde{CER}_{F}$. The minimum $\widetilde{CER}$ was 0.0706 and was achieved by the $SSDR_{Bod}$ method at $r = 2$ with a \textit{SD} of 0.0071. For $n_{i} = 6p = 300 < p^{2}/2$, unlike in Configurations 1-3,  for the \textit{SSDR}-based $\widetilde{CER}$, we expected a larger reduction in $\widehat{CER_{F}}$. The global minimum $\widetilde{CER}$ was roughly $0.0700$ for the $SSDR_{S^{-1}}$, $SSDR_{Haff}$, and $SSDR_{Wang}$  methods, all with $r = 2$. Moreover, the $SSDR_{Bod}$ and $SSDR_{MRY}$  techniques yielded the two smallest $\widetilde{CER}$s, both at $r = 2$ of 0.0646 and 0.0665, respectively. Thus, when the data dimensionality was reduced from $p = 50$ to $r = 2$, all five considered \textit{SSDR}-based approaches significantly reduced $\widetilde{CER}_{F}$ by approximately 0.0750. 

In the \textit{MC} simulation, the inclusion of the $SSDR_{S^{-1}}$ method demonstrated our proposed $SSDR$ technique's ability to preserve or primarily improve upon the $\widetilde{CER}_{F}$ in subspaces of dimension $r < p$. More importantly, our MC simulation illustrated a significant improvement in classifier performance achieved through the $SSDR$  approach, which incorporates a user-specified precision shrinkage estimator to stabilize the \textit{LDR} projection matrix. Across all of the \textit{MC} simulations, we determined that the $SSDR_{MRY}$ method provided superior or competitive $\widetilde{CER}$s among the five \textit{SSDR} methods considered. Moreover, the $SSDR_{MRY}$ method exhibited superior classifier stability, often with the smallest \textit{SD}, resulting in similar $\widetilde{CER}$s regardless of the class-specific training-sample size. The $SSDR_{Wang}$ and $SSDR_{Bod}$ methods performed similarly to the $SSDR_{MRY}$ method but yielded slightly larger $\widetilde{CER}$s for certain parameter and training-sample size scenarios.

\section{Real Data Applications}\label{sec:real_data}

\subsection{Data and Analysis Description}\label{subsec:data_sim_descr} 

\begin{table}[t!]
    \centering
    \caption{Datasets used for  contrasts that are described in Section \ref{subsec:data_sim_descr}. Class-specific sample sizes, denoted by $n_{i}$, are separated by colons.}. 
    \begin{tabular}{|c|cccc|}
    \hline
    Dataset & $k$ & $n$ & $n_{i}$ & $p$ \\
    \hline \hline
    Autism (AUT) & 2 & 98 & 36:62 & 18 \\
    Breast Cancer (BRC) & 2 & 683 & 444:239 & 9 \\
    Divorce (DIV) & 2 & 170 & 86:84 & 54 \\ 
    Ionosphere (ION) & 2 & 351 & 225:126 & 32 \\
    SPECT Heart (SPH) & 2 & 266 & 211:55 & 22 \\
    Penguins (PNG) & 3 & 333 & 146:68:119 & 6 \\ 
    Wheat Seeds (WHS) & 3 & 210 & 70:70:70 & 7 \\
    Ecoli (ECL) & 5 & 327 & 143:77:35:20:52 & 7 \\
    Dry Beans (BNS) & 7 & 13,611 & 2027:1322:522:1630:1928:2636:3546 & 16 \\
    \hline
    \end{tabular}
    \label{tbl:datasets}
\end{table}

\noindent Here, we considered nine real datasets from the University of California - Irvine (UCI) Machine Learning Repository. These are listed in Table \ref{tbl:datasets}.  The number of populations ranged from 2 to 7 classes, the total sample sizes varied from 98 to 13,611, and the number of original features ranged from 6 to 54. More information on the datasets is given in \ref{Appx:Data}. We remark that, when it was needed, we ``studentized" the data or  added random error from the $N(0, \sigma = 10^{-5})$ distribution to features causing covariance matrix singularity issues.

We used the nine real datasets given in Table \ref{tbl:datasets} to contrast the classification efficacy of the same five \textit{SSDR} methods considered in the \textit{MC} simulation in Section \ref{sec:mc_sim}. Again, we reduced the data dimension using each \textit{SSDR} method and then applied the \textit{QDA} classifier. For each dataset and \textit{SSDR} method, we performed repeated 10-fold cross-validation in which we took the mean value of the $\widehat{CER}$s for the 10 folds, denoted as $\overline{\textit{CER}}$, and then repeated this process 1,000 times. The classification performance was evaluated in terms of the median value of the $\overline{\textit{CER}}$s, $\widetilde{\textit{CER}}$, and we also calculated the \textit{SD} of the $\overline{\textit{CER}}$s. We then reported the $\widetilde{CER}_{F}$s, the $\widetilde{CER}$s for the reduced dimension, $r^{\ast}$, that achieved the global minimum $\widetilde{CER}$ for each \textit{SSDR} method, and the corresponding \textit{SD}s in Table \ref{tbl:SDR}. 

For the Breast Cancer, Divorce, Ionosphere, and Penguins datasets, we used the user-specified matrices proposed by \cite{molstad2018} for \textit{QDA} in the $SSDR_{MRY}$ method. Additionally, because of different magnitudes among the  features, we determined that standardizing the mean vectors in the recommended \textit{QDA}-specified matrices improved classification performance for the Breast Cancer, Ionosphere, and Penguins datasets. For the remaining datasets, we used the simple user-specified matrices for the $SSDR_{MRY}$ technique.

\subsection{Real Data Contrasts of Precision Matrix Estimators for SDR}\label{subsec:real_data_SSDR}

\noindent  In Table \ref{tbl:SDR}, for the Autism, Breast Cancer, Ionosphere, and Penguins datasets, we see that we obtained $\widetilde{CER}$ reductions for the chosen subspaces when they were contrasted to the full feature-dimensional $\widetilde{CER}_{F}$, regardless of the particular \textit{SSDR} method used.  In the table, we display only the global minimum $\widetilde{CER}$ corresponding to the selected reduced dimensions $r^{\ast} < p$. However, when $r = p$ held, our \textit{SSDR} method still preserved or improved on the $\widetilde{CER}_{F}$ for all datasets, regardless of the precision matrix estimator used. 
 
Notably, across all datasets considered here, the $SSDR_{MRY}$  technique consistently outperformed its five competitors by yielding the smallest $\widetilde{CER}$s, thus providing the largest reduction to the corresponding  $\widetilde{CER}_{F}$s. For example, for the Autism dataset, the $SSDR_{MRY}$ method achieved a $\widetilde{CER} = 0.0111$ compared to the $\widetilde{CER}_{F} = 0.1400$. Thus, for this dataset, the $SSDR_{MRY}$ method significantly reduced the $\widetilde{CER}_{F}$ by 0.1289, while reducing the dimensionality from $p = 18$ to $r = 2$. Additionally, for the Penguins dataset, the $SSDR_{MRY}$ method yielded an impressive $\widetilde{CER} = 0.00$. This result provided a reduction of $0.0810$ to the $\widetilde{CER}_{F}$. Moreover, the $SSDR_{MRY}$ method exhibited superior classifier stability for \textit{QDA} with the smallest \textit{SD} for seven of the nine datasets. We attribute the excellent performance of the $SSDR_{MRY}$ approach to the induced sparsity of the \textit{MRY} estimator. In fact, we determined that the percentage of zero entries in the averaged \textit{MRY} precision matrix estimators from our repeated cross-validation simulation was 31\%, compared to approximately 5\% for the other  precision shrinkage estimators considered here. 

The $SSDR_{Haff}$  method generally exhibited subpar performance, often yielding approximately the same or slightly larger $\widetilde{CER}$s than $SSDR_{S^{-1}}$. Although second to $SSDR_{MRY}$, the $SSDR_{Wang}$ method consistently showed significant reductions in $\widetilde{CER}$s across all datasets.  Additionaly, the $SSDR_{Wang}$ approach also had smaller $\widetilde{CER}$s when contrasted to the $SSDR_{S^{-1}}$ method and the $\widetilde{CER}_{F}$. Additionally, for the Divorce dataset, $SSDR_{Wang}$ achieved the same global minimum $\widetilde{CER}$ of 0.0111 as $SSDR_{MRY}$. The $SSDR_{Bod}$ $\widetilde{CER}$s notably improved the $\widetilde{CER}_{F}s$ for all datasets except the Dry Beans and Wheat Seeds datasets.  However, the $SSDR_{Bod}$ method was always inferior to the $SSDR_{MRY}$ approach.

\begin{table}[t!]
    \centering
    \caption{Median estimated conditional error rates, denoted  $\widetilde{CER}$, for real data classification applications described in Section \ref{subsec:data_sim_descr} for contrasting \textit{SSDR} methods coupled with \textit{QDA}. The global minimum $\widetilde{CER}$\ \% is given followed by the (\textit{SD}) and then the reduced-dimension size that achieves the minimum $\widetilde{CER}$\ \%, [$r^{\ast}$]. The full dimensional $\widetilde{CER}_{F}\ \%$ using the \textit{QDA} is denoted by \textit{QDA}. Smallest  observed $\widetilde{CER}$s are typeset in bold per dataset.}
    \begin{tabular}{|c|cccccc|}
        \hline
        Dataset & QDA & $SSDR_{S^{-1}}$ & $SSDR_{Haff}$ & $SSDR_{Wang}$ & $SSDR_{Bod}$ & $SSDR_{MRY}$ \\
        \hline \hline
        AUT  & 14.0 (2.02)   & 10.1 (1.77) [6]   & 11.4 (1.77) [5]   & 6.75 (1.31) [5]   & 5.02 (1.35) [4]   & \textbf{1.11} (0.75) [2] \\
        BRC  & 4.97 (0.15)   & \textbf{3.80} (0.19) [2]   & 4.10 (0.14) [3]   & \textbf{3.80} (0.14) [2]   & \textbf{3.80} (0.19) [2]   & \textbf{3.80} (0.14) [2] \\
        DIV  & 13.5 (0.61)   & 13.6 (0.48) [2]   & 13.6 (0.48) [2]   & \textbf{1.11} (0.47) [3] & 1.18 (0.21) [1]   & \textbf{1.11} (0.47) [3] \\
        ION  & 12.5 (0.49)   & 5.99 (0.71) [8]   & 5.99 (0.71) [8]   & 6.24 (0.53) [8]   & 6.27 (0.70) [8]   & \textbf{5.19} (0.43) [9] \\
        SPH  & 21.7 (1.03)   & 20.0 (0.85) [21]  & 19.9 (0.82) [18]  & 16.6 (0.93) [2]   & 16.5 (0.80) [1]   & \textbf{15.8} (0.56) [1] \\
        PNG  & 8.10 (0.52)   & 17.3 (0.98) [5]   & 17.3 (0.98) [5]   & 0.60 (0.36) [5]   & 0.29 (0.24) [4]   & \textbf{0.00} (0.19) [5] \\
        WHS  & 5.24 (0.41)   & 6.19 (0.65) [6]   & 6.19 (0.65) [6]   & 3.81 (0.45) [6]   & 6.19 (0.67) [6]   & \textbf{2.86} (0.49) [5] \\
        ECL  & 23.1 (0.96)   & 26.0 (1.07) [6]   & 25.4 (1.23) [6]   & 17.2 (1.05) [5]   & 18.9 (0.87) [6]   & \textbf{11.3} (0.56) [4] \\
        BNS  & 8.85 (0.04)   & 8.99 (0.04) [15]  & 8.99 (0.04) [15]  & 8.60 (0.04) [7]   & 8.99 (0.04) [15]  & \textbf{8.29} (0.04) [5] \\
        \hline
    \end{tabular}
    \label{tbl:SDR}
\end{table}

\subsection{Real Data Contrasts of Current SDR and Supervised Classification Methods}\label{subsec:sim_results}

\noindent In both the MC simulation and real data applications considered in Sections \ref{sec:mc_sim} and \ref{subsec:real_data_SSDR}, we determined that the $SSDR_{MRY}$ method produced a well-conditioned projection matrix due to the prominent sparsity in the \textit{MRY} precision matrix estimator.  Here, using the same nine real datasets as in Section \ref{subsec:real_data_SSDR}, we contrasted the performance of our newly-proposed $SSDR_{MRY}$ method with four current \textit{SDR} and supervised classification methods. These techniques consist of a sparse linear discriminant method (\textit{MSDA}) from \cite{mai2019}, an updated version of \textit{SIR} that uses lasso regression (\textit{SIRL}) proposed by \cite{lin2019}, the envelope discriminant subspace method (\textit{ENDS}) of \cite{zhang2019}, and a one-dimensional \textit{SDR} method that is based on the optimal one-dimensional \textit{QDA} error rate (\textit{QDAP}) from \cite{wu2022}. We remark that the approach derived by \cite{wu2022} was applicable for only two populations. 

\begin{figure}[t!]
    \centering
    \includegraphics{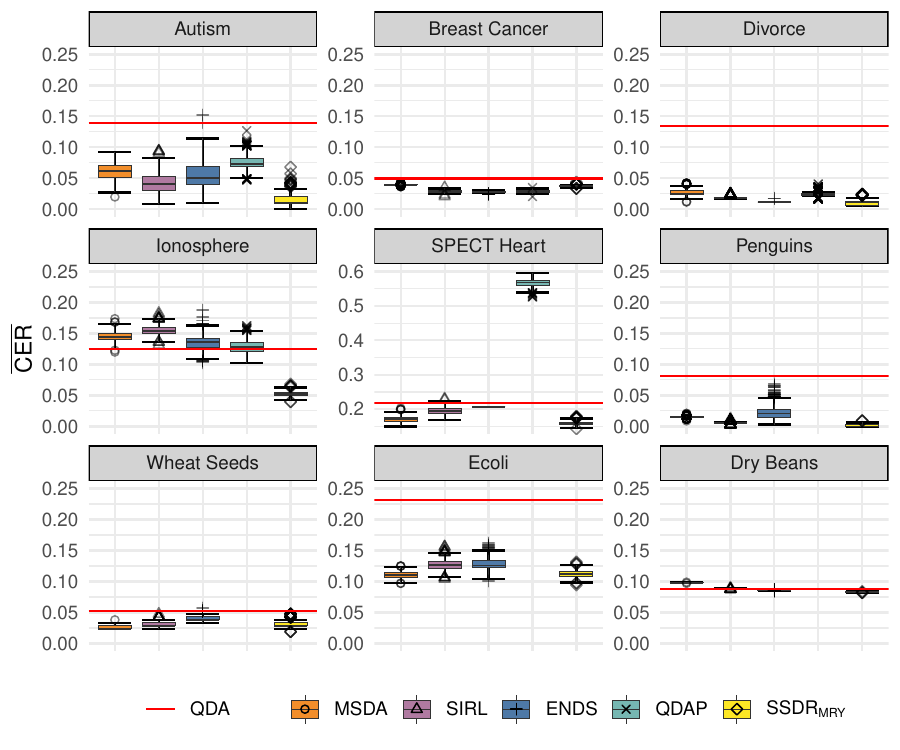}
    \caption{Mean 10-fold cross-validation \textit{conditional error rate}, denoted $\overline{CER}$, plots for comparison of \textit{MSDA}, \textit{SIRL}, \textit{ENDS}, \textit{QDAP}, and $SSDR_{MRY}$ methods as described in Section \ref{subsec:sim_results}. The full-dimensional median $\overline{CER}$ using  \textit{QDA} is given by the horizontal line. The \textit{QDAP} method is  applicable only to binary response data and is not displayed for some datasets. The y-axis scale is from 0.00 to 0.25 for all datasets except the SPECT Heart dataset.} 
    \label{fig:2}
\end{figure}

To compare the performance of each method, we performed the repeated 10-fold cross-validation procedure described in Section \ref{subsec:data_sim_descr} to obtain estimates for the \textit{CER}. For fair comparison, we used \textit{QDA} as the classifier for each \textit{SDR}-based method. We have graphically displayed the $\overline{\textit{CER}}s$ for each method in Figure \ref{fig:2}. When applicable, to minimize the corresponding $\widetilde{CER}$s, we determined all reduced dimensions and tuning parameters via cross-validation for each method.

The proposed $SSDR_{MRY}$ method outperformed the competing \textit{ENDS}, \textit{SIRL}, \textit{QDAP}, and \textit{MSDA} methods by achieving the smallest $\widetilde{CER}$ for six  of the nine datasets. These six datasets were the Autism, Divorce, Ionosphere, SPECT Heart, Penguins, and Dry Beans datasets. Additionally, our $SSDR_{MRY}$ technique was the only method that consistently improved upon the $\widetilde{CER}_{F}$ for every dataset. For example, for the Ionosphere dataset, with the exception of our $SSDR_{MRY}$ method, all the \textit{SDR} and classification methods yielded $\widetilde{CER}$s greater than the $\widetilde{CER}_{F} = 0.1250$. However, for the Ionosphere dataset, our proposed $SSDR_{MRY}$ method reduced the data dimension from $p = 32$ to $r = 8$ and achieved a $\widetilde{CER}$ = 0.0519. Thus, $SSDR_{MRY}$ reduced the $\widetilde{CER}_{F}$ by 0.0731.  Once again, we attribute the $SSDR_{MRY}$ method's efficacy to its preservation of differences in the classes and the improved stability of its projection matrix, which prevents over-parameterizing regardless of the training-sample sizes.

The \textit{ENDS} method achieved the minimum $\widetilde{CER} = 0.0279$ for the Breast Cancer dataset; however, this $\widetilde{CER}$ was within a standard error of the $\widetilde{CER}$s of the \textit{QDAP} and \textit{SIRL} methods. The \textit{SIRL} method yielded competitive $\widetilde{CER}$s. However, the method never achieved the smallest $\widetilde{CER}$ for any of the nine datasets. Moreover, the \textit{SIRL} method yielded a $\widetilde{CER}$ greater than the $\widetilde{CER}_{F}$ for the Dry Beans and Ionosphere datasets. We note that the \textit{ENDS} and \textit{SIRL} methods often achieved their minimum $\widetilde{CER}$ in the same or smaller subspace dimension as the $SSDR_{MRY}$ method. However, we believe that their bias towards smaller reduced dimensions led to a lack of robustness for datasets with a large original feature-space, often resulting in larger $\widetilde{CER}$s because of a reduced data representation that was too small to preserve all important original discrimination information. The \textit{QDAP} performance was similar to that of the considered competitive methods, except for the SPECT Heart dataset, where it yielded a significantly larger $\widetilde{CER} = 0.5678$ when contrasted to the $\widehat{CER_{F}} =  0.2170$. The \textit{MSDA} method achieved the minimum $\widetilde{CER}$s of 0.0238 and 0.1106 for the Wheat Seeds and Ecoli datasets, respectively. However, for these two datasets, the $SSDR_{MRY}$  method gave $\widetilde{CER}$s of 0.0286 and 0.1127, which were only slightly greater than the \textit{MSDA} $\widetilde{CER}$s. Additionally, similar to the \textit{SIRL} method, for the Dry Beans dataset, the \textit{MSDA} method had a larger $\widetilde{CER}$ of 0.0979 when contrasted to the $\widetilde{CER}_{F}$ = 0.0885. 

\section{Discussion}\label{sec: discussion}
\noindent In this paper, we derived a theoretical \textit{SDR} method that preserves the optimal error rate under the \textit{QDA} decision rule in a reduced dimension, $q < p$. Additionally, when parameter estimation is required, we have proposed a new \textit{SDR} technique, \textit{stabilized SDR} (SSDR), which utilizes a user-specified precision matrix shrinkage estimator to improve stability of the projection matrix. This proposed \textit{SSDR} method is particularly useful, though not limited to, training-sample-size scenarios where $p < n_i < p^2/2$. We have examined multiple precision matrix estimators to use in our proposed \textit{SSDR} method. Through \textit{MC} simulations and real data applications, we have determined that the $SSDR_{MRY}$ method improved the dimension reduction subspace for supervised classification, often preserving class differences more effectively than the other precision matrix estimators considered here. Additionally, the $SSDR_{MRY}$ method induces sparsity, which reduces the number of parameters to be estimated and significantly decreases the variability of the \textit{QDA} classifier.

Through empirical studies, we have demonstrated that our $SSDR_{MRY}$ method often achieved superior classification efficacy by producing smaller estimated \textit{CER}s, thus  frequently outperforming several current competitive \textit{SDR} and supervised classification methods. Although we do not claim that $SSDR_{MRY}$ always yields the smallest estimated \textit{CER}, we have found considerable evidence that when training-sample sizes are small relative to the original feature-space dimension, the $SSDR_{MRY}$ method in conjunction with the \textit{QDA} decision rule offers a highly competitive \textit{SDR} and classification method for heteroscedastic populations. In summary, we have developed a new \textit{SDR} method that provides improved projection matrix stability and, therefore, \textit{QDA} classifier stability. We have promoted \textit{SDR}-enhancing properties through induced sparsity and improved precision matrix estimation by employing a user-specified precision matrix shrinkage estimator.

\subsection{Code Availability}
\noindent We performed the MC simulations  and real data applications using the statistical programming language \texttt{R}, version 4.3.1.  We implemented \texttt{R} package \texttt{HDShOP} for the \textit{Bodnar} estimator and the \texttt{R} package \texttt{SCPME} for the \textit{MRY} precision matrix estimator. In addition, we programmed the \textit{Haff} and \textit{Wang} precision matrix estimators because they were not readily available in \texttt{R}. We implemented  the  \textit{MSDA} and \textit{SIRL} approaches in the \texttt{R} packages \texttt{TULIP} and \texttt{LassoSIR}, respectively, and  also formatted the \textit{ENDS} method in MATLAB using the code provided by the original authors in their supplementary material. Additionally, we configured the \textit{QDAP} method using the \texttt{R} package available on the authors' \href{https://github.com/ywwry66/QDA-by-Projection-R-Package}{github} site. Our new \textit{SSDR} method is implemented in the working \texttt{sdr} package available at \url{https://github.com/D3r1kBoonstra/sdr}. Lastly, one can find all reproducible code for the simulations and graphical figures at \url{https://github.com/D3r1kBoonstra/Stabilized_SDR}.

\section*{Acknowledgements} 
\noindent The authors of this research article received no grant money from funding agencies in the public, commercial, or not-for-profit sectors. We also thank Mrs. Joy Young for her grammar recommendations that enhanced the quality of our writing.

\appendix
\renewcommand{\thesection}{Appendix \Alph{section}}

\hypersetup{bookmarksdepth=-1}
\section{Additional Dataset Information}\label{Appx:Data}
\noindent Here, we give additional information on the datasets used in Section \ref{subsec:sim_results}. The Autism data from \cite{tabtah2017} pertains to autistic spectrum disorder screening for adolescents. The Wisconsin Breast Cancer Diagnostic data is from \cite{street1993} and contains data to classify  breast tissue as benign or malignant using features from images of a fine needle aspirate of the breast tissue. The Divorce data, introduced by \cite{yontem2019}, consists of an ordinal questionnaire, referred to as the Divorce Predictor Scale. The Ionosphere data originates from \cite{sigillito1989} and classifies electrons as ``good'' or  ``bad'' using features from a signal collection system. The SPECT Heart data from \cite{kurgan2001} covers abnormalities in cardiac single-proton emission-computed tomography. Palmer's Penguins data from \cite{gorman2014} is likened to the Iris data in \cite{fisher1936} for teaching machine learning and uses various measurements of penguins to classify them into particular species. The Wheat Seed data, provided by \cite{charytanowicz2010}, consists of geometric parameters from X-ray images of three wheat kernel species. The Ecoli data from \cite{horton1996} refers to the cellular location sites of protein. We removed classes omL, imL, and imS because of sample covariance matrix singularity due to insufficient class-specific training-sample sizes. Lastly, extracted features from high-resolution camera images were used to create the Dry Beans data in \cite{koklu2020}.

\hypersetup{bookmarksdepth=-1}
\section{Proofs of Three Lemmas}\label{Appx:Lemmas}

\begin{lemma}\label{lemma1}
Let $\bf{W} \in \R^{p \times s}$, where $s \coloneqq (k - 1)(p + 1),$ be
\begin{equation*}
\bf{W} \coloneqq \left[ \bf{g}_2 - \bf{g}_1 |\ldots|\bf{g}_k - \bf{g}_1 |\bf{H}_2 - \bf{H}_1 |\ldots|\bf{H}_k - \bf{H}_1 \right],
\end{equation*}
where $\bf{g}_i \in \R^{p}$, $\bf{H}_i \in \mathbb{S}^p$, and $\bf{g}_1 \ne \bf{g}_j$ and $\bf{H}_1 \ne \bf{H}_j$ for at least one value of $k$, where $2 \le j \le k$ and $i = 1, \ldots, k$.  Also, let rank$(\bf{W}) = 1 \le q < p$, and let $\bf{U} \in \R^{p \times q}$, $\bf{U}^{T}\bf{U} = \bf{I}_{q}$, $\bf{D} \in \R^{q \times q}$, and $\bf{V} \in \R^{s \times q}$ be the matrix components of the \textit{reduced SVD} of $\bf{W}$ such that $\bf{W} = \bf{UDV}^{T}$ with rank$(\bf{U}) = q$, and let ${\bf{P}}_{\bf{U}} \coloneqq {\bf{UU}}^{T}$. Then,
\begin{enumerate}[(a)]
\item ${\bf{P}}_{\bf{U}}\left(\bf{g}_i - \bf{g}_1\right) = \bf{g}_i - \bf{g}_1$; ${\bf{P}}_{\bf{U}}\left(\bf{H}_i - \bf{H}_1\right) = \bf{H}_i - \bf{H}_1$,
\item ${\bf{P}}^{\perp}_{{\bf{U}}}\left(\bf{g}_i - \bf{g}_1\right) = \bf{0}$; ${\bf{P}}^{\perp}_{{\bf{U}}}\left(\bf{H}_i - \bf{H}_1\right) = \bf{0}$,
\item ${\bf{P}}_{\bf{U}}\left(\bf{H}_i - \bf{H}_1\right) = \left(\bf{H}_i - \bf{H}_1\right){\bf{P}}_{\bf{U}}$,
\item ${\bf{P}}_{\bf{U}}\bf{H}_i = \bf{H}_i{\bf{P}}_{\bf{U}}$, 
\item ${\bf{P}}^{\perp}_{{\bf{U}}}\bf{H}_i = \bf{H}_1{\bf{P}}^{\perp}_{{\bf{U}}}$.
\end{enumerate}

\begin{proof}
For parts (a) and (b), note that $\bf{g}_i - \bf{g}_1$,  $\bf{H}_i - \bf{H}_1 \in \mathcal{C}(\bf{U})$, and $\bf{g}_i - \bf{g}_1$,  $\bf{H}_i - \bf{H}_1 \in \mathcal{N}(P^{\perp}_{{\bf{U}}})$, where $i \in \left\{2,...,k\right\}$. For part (c), because $\bf{H}_i\in\mathbb{S}^p$, we have
\begin{align*}
{\bf{P}}_{{\bf{U}}}\left[\bf{H}_i - \bf{H}_1\right] = \left[\bf{H}_i - \bf{H}_1\right]^T =  \left[\bf{H}_i - \bf{H}_1\right]{\bf{P}}_{{\bf{U}}}.
\end{align*}
For part (d), recall that for $\bf{x}\in\R^{p\times 1}$, $\bf{x}^TP_{{\bf{U}}}$ projects $\bf{x}$ onto the row space of ${\bf{P}}_{{\bf{U}}}$. Because ${\bf{P}}_{{\bf{U}}}\in\mathbb{S}_p$, the column space and row space are equal. Thus, ${\bf{P}}_{{\bf{U}}}\bf{H}_i = \bf{H}_i{\bf{P}}_{{\bf{U}}}$. Finally, for (e), we have by parts (b) and (d) that
\begin{align*}
{\bf{P}}^{\perp}_{{\bf{U}}}\left(\bf{H}_i - \bf{H}_1\right) = \bf{0} \quad
& \Longleftrightarrow \quad {\bf{P}}^{\perp}_{{\bf{U}}}\bf{H}_i = \bf{H}_1 - {\bf{P}}_{{\bf{U}}}\bf{H}_1 \\
&\Longleftrightarrow \quad {\bf{P}}^{\perp}_{{\bf{U}}}\bf{H}_i = \bf{H}_1 - \bf{H}_1{\bf{P}}_{\bf{U}} \\
&\Longleftrightarrow \quad {\bf{P}}^{\perp}_{{\bf{U}}}\bf{H}_i = \bf{H}_1{\bf{P}}^{\perp}_{{\bf{U}}}.
\end{align*}
\end{proof}
\end{lemma}

\begin{lemma}\label{lemma2}
    Consider the matrices $\bf{U}$, $\bf{P}_{\bf{U}}$, and $\bf{H}_i$, $i = 1, \ldots, k$, defined in  Lemma \ref{lemma1}. Then, $\left[ \bf{U}^{T}\bf{H}_i\bf{U} \right]^{-1} = \bf{U}^T\bf{H}^{-1}_i\bf{U}$.
 
\begin{proof}
Using Lemma \ref{lemma1}(d), we have that 
\begin{align*}
(\bf{U}^{T}\bf{H}_{i}\bf{U}) (\bf{U}^{T}\bf{H}_{i}^{-1}\bf{U}) &= \bf{U}^{T}\bf{H}_{i}\bf{P}_{\bf{U}}\bf{H}_{i}^{-1}\bf{U} \\
&= \bf{U}^{T}\bf{P}_{\bf{U}}\bf{H}_{i}\bf{H}_{i}^{-1}\bf{U}\\
&= \bf{U}^{T}\bf{U}\\
&= \bf{I}.
\end{align*}
\end{proof}
\end{lemma}

\begin{lemma}\label{lemma3}
 Consider $\bf{U}$, $\bf{g}_i$,  and $\bf{H}_i$, $i = 1, \ldots, k$, defined in  Lemma \ref{lemma1}. Also, let $\bf{C} =  \bf{R}\left[\bf{I} - \bf{UU}^{T}\right] \in \R^{(p - q) \times p}$, where $\bf{R} \in \R^{(p - q) \times p}$ such that rank$(\bf{C}) = p - q$. Then, we have that
 \begin{enumerate}[(a)]
 \item $\bf{Cg}_i = \bf{Cg}_1,$
   \item $\bf{CH}_i\bf{C}^T = \bf{CH}_1 \bf{C}^T$.
 \end{enumerate}

\begin{proof}
The proof of (a) follows trivially from Lemma \ref{lemma1}(b). Also, by Lemma \ref{lemma1}(d), note that
\begin{align*}
\bf{U}^{T}\bf{H}_i\bf{C}^T &= \bf{U}^{T}\bf{H}_i\left(\bf{I}_p - \bf{UU}^{T}\right)^T\bf{R}^T \\
&= \left(\bf{U}^{T}\bf{H}_i - \bf{U}^{T}\bf{H}_i\bf{UU}^{T}\right)\bf{R}^T \\
&= \bf{0}.
\end{align*}
Then by Lemma \ref{lemma1} (b), $\bf{CH}_i\bf{C}^T - \bf{CH}_i\bf{U}\left(\bf{U}^T \bf{H}_i\bf{U}\right)^{-1}\bf{U}^T \bf{H}_i\bf{C}^T = \bf{CH}_1\bf{C}^T$.
\end{proof}
\end{lemma}

\bibliographystyle{cas-model2-names}
\bibliography{main.bbl}

\end{document}